\documentclass[onecolumn,  
showpacs,  
preprintnumbers,  
aps,  
prd,  
a4paper,  
nofootinbib,  
tightenlines,  
floats, floatfix  
]{revtex4}

\usepackage{amsmath}
\usepackage{amsfonts}
\usepackage{graphicx,color}

\newcommand{\bea}{\begin{eqnarray}}
\newcommand{\ena}{\end{eqnarray}}

\newcommand{\RR}{{\cal R}}

\newcommand{\PP}{{\cal P}}

\renewcommand{\a}{\alpha}

\renewcommand{\d}{\delta}
\newcommand{\D}{\Delta}

\newcommand{\e}{\epsilon}

\renewcommand{\l}{\lambda}

\begin{document}


\title{Inflation coupled to a Gauss-Bonnet term}

\author{Peng-Xu Jiang$^{1,2,}$\footnote{Email: shuiruge@gmail.com}}
\author{Jian-Wei Hu$^{2,}$\footnote{Email: jwhu@itp.ac.cn}}
\author{Zong-Kuan Guo$^{2,}$\footnote{Email: guozk@itp.ac.cn}}

\affiliation{$^1$Department of Astronomy, Beijing Normal University, Beijing 100875, China}
\affiliation{$^2$State Key Laboratory of Theoretical Physics, Institute of Theoretical Physics,
Chinese Academy of Sciences, P.O. Box 2735, Beijing 100190, China}





\date{\today}

\begin{abstract}
The newly released Planck CMB data place tight constraints on
slow-roll inflationary models. Some of commonly discussed
inflationary potentials are disfavored due mainly to the large
tensor-to-scalar ratio. In this paper we show that these potentials
may be in good agreement with the Planck data when the inflaton has
a non-minimal coupling to the Gauss-Bonnet term.
Moreover, such a coupling violates the consistency relation between
the tensor spectral index and tensor-to-scalar ratio.
If the tensor spectral index is allowed to vary freely,
the Planck constraints on the tensor-to-scalar ratio are slightly improved.
\end{abstract}


\maketitle

\section{Introduction}

Inflation in the early Universe not only provides a way to solve the
flatness and horizon problems of the standard big bang cosmology
but also produces density perturbations as seeds for large-scale
structure in the Universe. The simplest scenario of cosmological
inflation is based upon a single, canonical and minimally-coupled
scalar field with a flat potential. In this scenario, quantum
fluctuations of the inflaton give rise to an almost scale-invariant,
nearly Gaussian and adiabatic power spectrum of curvature perturbations.
This prediction can be directly tested by the measurement of the
temperature anisotropies in cosmic microwave background (CMB).
The latest CMB data, from the Planck satellite with its combination
of high sensitivity, wide frequency range and all-sky coverage,
have placed strong constraints on inflationary models
from the information of the spectral index of curvature
perturbations as well as the tensor-to-scalar ratio~\cite{hin12,ade13}.

String theory is currently regarded as the most promising candidate
for unifying gravity with the other fundamental forces and for
a theory of quantum gravity. Typically there are correction terms
of higher orders in the curvature to the lowest order effective
supergravity action coming from superstrings, which may play a
significant role in the early Universe. Interestingly, there is
a unique combination of the curvature squared terms, the Gauss-Bonnet (GB) term,
which is ghost-free in Minkowski background and keeps the order of
the gravitational equations of motion unchanged.
Such a term appears in the tree-level effective action of the heterotic string~\cite{cal85}.
At the one-loop level of string effective action it may arise as
the Gauss-Bonnet term coupled to a modulus field, which provides
the possibility of avoiding the initial singularity of the Universe~\cite{ant92,kaw98,top02}.
There have been many works that discuss accelerating cosmology
with the GB coupling term in four and higher dimensions~\cite{guo07a,noj05,bro05}.
Moreover, the effect of the GB coupling term on the evolution of primordial
perturbations was investigated in~\cite{hwa99,sat08,guo07}.

It is known that the GB term coupled to the scalar field can
drive inflation as the effective potential, which, however, gives rise
to violent negative instabilities of tensor perturbations around
a de Sitter background on small scales~\cite{guo07}.
In power-law inflation implemented by an exponential potential and
an exponential GB coupling, tensor or scalar perturbations
exhibit negative instabilities on small scales when the GB coupling
dominates the dynamics of the background~\cite{guo09}.
In such a model when the potential is dominant, the GB coupling term with
a positive (or negative) coupling may lead to a reduction
(or enhancement) of the tensor-to-scalar ratio.
The more general formalism of slow-roll inflation with an arbitrary
potential and an arbitrary coupling was developed by introducing
a combined hierarchy of Hubble and GB flow functions~\cite{guo10}.
In this scenario the standard consistency relation between the
tensor-to-scalar ratio and the spectral index of tensor perturbations
does not hold.

In this paper, we apply the general formalism developed
in~\cite{guo10} to some specific inflationary models.
To confront the models with observational data, we use recent
CMB measurements by the Planck experiment to bound on the
tensor-to-scalar ratio and scalar spectral index when
the tensor spectral index is allowed to vary freely.
We find that the GB coupling may effectively suppress the
tensor-to-scalar ratio, which can improve the fit to the data.

The organization of the paper is as follows. In Section~\ref{sec2}
we outline the relevant features of slow-roll inflation with
an inflaton coupled non-minimally to the GB term.
In Section~\ref{sec3} we confront some of commonly discussed
inflationary models with the Planck data.
Section~\ref{sec4} is devoted to discussions and conclusions.

\section{slow-roll inflation}
\label{sec2}

We consider the following action
\bea
S= \int d^4 x \sqrt{-g} \left[\frac{R}{2} - \frac12\partial_{\mu} \phi \partial^{\mu} \phi - V(\phi) - \frac12\xi(\phi)R^2_{\rm GB}\right],
\label{act}
\ena
where $\phi$ is the inflaton field with a potential $V(\phi)$,
$R$ is the Ricci scalar, $R^2_{\rm GB} \equiv R_{\mu\nu\rho\sigma}R^{\mu\nu\rho\sigma} - 4R_{\mu\nu}R^{\mu\nu} + R^2$
is the GB term, and $\xi(\phi)$ is a coupling function of $\phi$.
We work in Planckian units, i.e., $\hbar=c=8\pi G=1$.
In the weak coupling limit of the low-energy effective string
theory, the coupling may take the form of $\xi \propto e^{-\phi}$~\cite{ant92}.
As the system enters a large coupling region, it is expected
that the form of the function $\xi(\phi)$ becomes complicated.
The potential may arise naturally from supersymmetry breaking or
other nonperturbative effects. Hence, we work on the general action~(\ref{act}).
In a spatially flat Friedmann-Robertson-Walker Universe with the
scale factor $a$, from the action~(\ref{act}) we obtain the
background equations
\bea
\label{bge1}
&& 3H^2 = \frac12\dot{\phi}^2+V+12\dot{\xi}H^3 \;, \\
&& \ddot{\phi}+3H\dot{\phi} = - V_{,\phi} - 12\xi_{,\phi}H^2\left(\dot{H}+H^2\right) \;,
\label{bge2}
\ena
where a dot represents the time derivative, $(...)_{,\phi}$
denotes a derivative with respect to $\phi$, and $H\equiv \dot{a}/a$
is the Hubble parameter. Note that the coupling function $\xi$ works
as the effective potential for the inflaton $\phi$.

As discussed in~\cite{guo10}, since the new degree of freedom
is introduced by the GB coupling function $\xi(\phi)$, it is
useful to introduce a combined hierarchy of Hubble and Gauss-Bonnet
flow parameters. Following Refs.~\cite{sch01}, we define the
hierarchy as $\e_1=-\dot{H}/H^2$, $\d_1=4\dot{\xi}H$,
$\e_{i+1}=d\ln|\e_i|/d\ln a$ and $\d_{i+1}=d\ln |\d_i|/d\ln a$
for $i\ge 1$. The slow-roll conditions become $|\e_i|\ll 1$ and
$|\d_i|\ll 1$, analogous to the standard slow-roll approximation.
Under such conditions the background equations~(\ref{bge1})
and~(\ref{bge2}) reduce to
\bea
&& H^2 \simeq \frac13 V \;, \\
&& H\dot{\phi} \simeq -\frac13 V Q,
\ena
with $Q\equiv V_{,\phi}/V+4\xi_{,\phi}V/3$. If $Q=0$, the motion of
inflaton is frozen because of the force due to the slope of the
potential is exactly balanced by one from the GB coupling.
In the case of $V_{,\phi}\xi_{,\phi}>0$, the GB coupling makes
the evolution of the inflaton faster than in the case of standard
slow-roll inflation, which decreases the Hubble expansion rate.
If $V_{,\phi}\xi_{,\phi}<0$, since the GB coupling slows the
field evolution, inflation may occur even for a steep potential.
The number of e-folds is computed as the following
\bea
N(\phi) \simeq \int^{\phi}_{\phi_{\rm end}}\frac{d\phi}{Q(\phi)}\,.
\ena

The primordial power spectra of scalar and tensor perturbations
are derived in~\cite{guo10}
\bea
\PP_{\RR}&=&\frac{H^2}{4\pi^2 c^3_{\RR} F_{\RR}}\,, \\
\PP_{T}&=&\frac{2H^2}{\pi^2 c^3_{T} F_{T}}\,,
\ena
where the expressions are evaluated at the time of horizon crossing
at $c_{\RR}k=aH$ and $c_T k=aH$, respectively.
As shown in~\cite{guo10}, to lowest order in the slow-roll parameters
this difference of horizon-crossing time is unimportant.
We have assumed that time derivatives of the flow parameters
can be neglected during slow-roll inflation, which allows us
to obtain the leading contribution to the slow-roll approximation.
Here $c_{\RR}$, $F_{\RR}$, $c_{T}$ and $F_{T}$ are given by
\bea
\label{c2RR}
c^2_{\RR} &=& 1+\frac{8\D\dot{\xi}H\dot{H}+2\D^2H^2(\ddot{\xi}-\dot{\xi}H)}{\dot{\phi}^2+6\D\dot{\xi}H^3}\,,\\
F_{\RR} &=& \frac{\dot{\phi}^2+6\D\dot{\xi}H^3}{(1-\D/2)^2H^2}\,,\\
c^2_{T} &=& 1-\frac{4(\ddot{\xi}-\dot{\xi}H)}{1-4\dot{\xi}H}\,,\\
F_{T} &=& 1-4\dot{\xi}H,
\ena
with $\D\equiv4\dot{\xi}H/(1-4\dot{\xi}H)$. The tensor-to-scalar
ratio $r\equiv \PP_T/\PP_{\RR}$ and spectral indices of scalar
and tensor perturbations are given in terms of the Hubble and GB flow parameters
\bea
r &\simeq& 8\left(2\e_1-\d_1\right), \label{srr} \\
n_{\RR}-1 &\simeq& -2\e_1-\frac{2\e_1\e_2-\d_1\d_2}{2\e_1-\d_1}, \label{srn}\\
n_{T} &\simeq& -2\e_1.
\ena
For a positive $\e_1$, $QV_{,\phi}>0$ is required. In this scenario,
we see that the degeneracy of standard consistency relation between $r$ and
$n_T$ is broken due to the presence of the extra degree of freedom $\d_1$.
For this reason, the future experimental checking of this relation
is usually regarded as an important test of the simplest forms of inflation~\cite{che12}.
The tensor-to-scalar ratio is suppressed for a positive $\d_1$
while it is enhanced for a negative $\d_1$.
The Hubble and GB flow parameters can be expressed in terms of
the potential and GB coupling function
\bea
\e_1 &=& \frac{Q}{2} \frac{V_{,\phi}}{V} \label{equ:fp.1} \;, \\
\e_2 &=& -Q\left(\frac{V_{,\phi\phi}}{V_{,\phi}}-\frac{V_{,\phi}}{V}+\frac{Q_{,\phi}}{Q}\right) \label{equ:fp.2} \;, \\
\d_1 &=& -\frac{4Q}{3} \xi_{,\phi}V \label{equ:fp.3} \;, \\
\d_2 &=& -Q\left(\frac{\xi_{,\phi\phi}}{\xi_{,\phi}}+\frac{V_{,\phi}}{V}+\frac{Q_{,\phi}}{Q}\right) \label{equ:fp.4} \;.
\ena

\section{models and observations}
\label{sec3}

In this section we will study several inflationary models as an
illustration. We assume that the power spectra of scalar and tensor perturbations
can be parameterized as power-law at the pivot scale $k_0=0.002$
Mpc$^{-1}$. As described in the previous section, the inflation
consistency relation $n_T=-r/8$ is violated by the GB coupling.
Hence, $n_T$ should be allowed to vary independent of the
tensor-to-scalar ratio. We adopt a flat prior on $n_T$ of $[-3, 0]$.
We use the Planck CMB temperature likelihood~\cite{ade13}, which combines a Gaussian
likelihood approximation at high multipoles with a pixel-based
approach at low multipoles, supplemented by the large scale 9-year
WMAP polarization data~\cite{hin12} that gives a constraint on the
reionization optical depth. Figures~\ref{figB} and \ref{figC} show the Planck+WP
constraints in the $n_\RR-r$ plane for a varying $n_T$.
Compared to the Planck's results for the standard slow-roll
inflation~\cite{ade13}, relaxing the consistency relation leads
to a slightly tighter upper bound on $r<0.10$ at 95\% confidence level.
In the standard slow-roll inflation, the consistency relation imposes
a nearly scale-invariant spectrum of tensor modes since the upper
limits of $r$ is of the order $10^{-1}$. Deviations from scale
invariance lead to more contribution of tensor modes to the temperature power spectrum of the CMB.
In what follows we consider several commonly discussed inflationary
models in light of the Planck observations.
Comparisons of these models with observations are implemented
by using predictions of $n_\RR$ and $r$.

\subsection{power-law inflation}

Let us first consider power-law inflation with an exponential
potential and an exponential GB coupling
\bea
V(\phi)=V_0 e^{-\l \phi}, \quad \xi(\phi)=\xi_0 e^{\l \phi},
\ena
where $V_0$, $\xi_0$ and $\l$ are constants. For later
convenience we define $\a\equiv 4 V_0\xi_0/3$ throughout
the rest of the paper.
Such a model was considered as power-law inflation in~\cite{guo09}.
It can also provide an alternative explanation for the current
acceleration of the Universe~\cite{noj05}.
Like the standard power-law inflation, there is no natural
end to inflation within the model. Hence an additional
mechanism is required to stop it.
In the general slow-roll formalism this model, in which $\e_1=\l^2(1-\a)/2$, $\d_1=\l^2\a(1-\a)$,
and both $\e_2$ and $\d_2$ vanish, predicts $n_\RR-1=-\l^2(1-\a)$
and $r=8\l^2(1-\a)^2$. One gets the relation between $r$
and $n_\RR$ as
\bea
r=-8(1-\a)(n_\RR-1),
\ena
which indicates that a positive (or negative) $\a$ can suppress
(or enhance) the tensor-to-scalar ratio. It is known that
the model with $\a=0$ is now outside the joint $99.7\%$ CL region
in the $n_\RR-r$ plane derived from the Planck+WP data.
If $\a \gtrsim 0.5$, this class of models can be consistent with
the Planck constraints.


\subsection{chaotic inflation with an inverse power-law coupling}

\begin{figure}
\includegraphics[width=12cm,height=8cm]{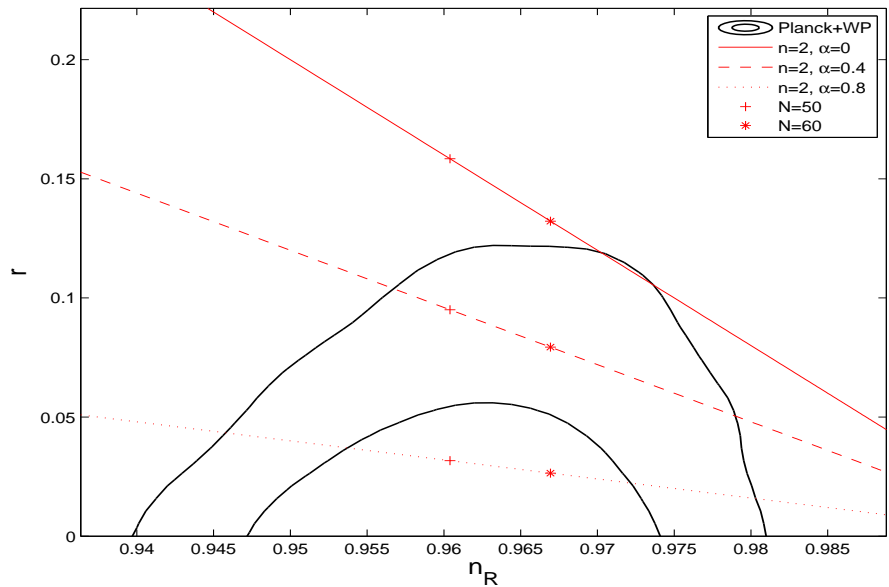}
\caption{Marginalized joint 68\% and 95\% CL regions for
$n_\RR$ and $r$ without the consistency relation from the Planck+WP data,
compared to the theoretical predictions of the model~\eqref{modelB}
with $n=2$.}
\label{figB}
\end{figure}

We now consider another model with a monomial potential and an
inverse monomial GB coupling
\bea
V(\phi)=V_0 \phi^n, \quad \xi(\phi)=\xi_0 \phi^{-n}.
\label{modelB}
\ena
This class of potentials has been widely studied as the simplest
inflationary model. Without the GB coupling the $\phi^4$ potential
is well outside of the joint $99.7\%$ CL region and the $\phi^2$
potential lies outside the joint $95\%$ CL for the Planck+WP+high-$l$
data, as discussed in~\cite{ade13}. Here we show that the GB
coupling may revive this class of potentials. Such a specific
choice of GB coupling allows us to find an analytic relation
between $r$ and $n_\RR$ in terms of $N$. This model, in which $\e_1=n/(4N+n)$,
$\d_1=2n\a/(4N+n)$ and $\e_2=\d_2=4/(4N+n)$, predicts~\cite{guo10}
\bea
n_\RR-1 &=& -\frac{2(n+2)}{4N+n},\\
r &=& \frac{16n(1-\a)}{4N+n}.
\ena
To confront the theoretical predictions with the Planck constraints,
in figure~\ref{figB} we plot the values of $n_\RR$ and $r$ in the
model with $n=2$ for different values of $N$ and $\a$. We see
that for a fixed number of e-folds the parameter $\a$ can shift
the predicted $r$ vertically and keep $n_\RR$ invariant.
Figure~\ref{figB} shows that the model with a positive $\a$ can
be consistent with the Planck data while one with a negative $\a$
is disfavored.


\subsection{chaotic inflation with a dilaton-like coupling}

In the two classes of models discussed above, we notice that the GB coupling
and potential satisfy $\xi(\phi)V(\phi)=3\a/4$, so that the
relation between $r$ and $n_\RR$ can analytically be expressed
in terms of model parameters. Here we consider a more general
model with a monomial potential and a dilaton-like coupling
\bea
V(\phi)=V_0 \phi^n \;, \quad \xi(\phi)=\xi_0 e^{-\l\phi}.
\label{modelC}
\ena
For this model the Hubble and GB flow parameters are given by
\begin{align}
  \epsilon_1 & = \frac{n \left(n-\alpha  \lambda  e^{-\lambda  \phi } \phi ^{n+1}\right)}{2 \phi ^2} \;, \\
  \epsilon_2 & = \frac{ 2 n -\alpha  \lambda e^{-\lambda  \phi } \phi ^{n+1} (\lambda  \phi-n+1)}{\phi ^2} \;, \\
  \delta_1 & = \frac {\alpha  \lambda  e^{- \lambda  \phi } \phi ^{n+1} \left(n -\alpha  \lambda e^{- \lambda  \phi } \phi ^{n+1}\right)}{\phi^2} \;, \\
  \delta_2 & = \frac{ n (\lambda  \phi - n +1) - 2 \alpha  \lambda  e^{-\lambda  \phi } \phi ^{n+1} (\lambda \phi -n)}{\phi ^2} \;.
\end{align}
From Eqs.~\eqref{srr} and \eqref{srn} one gets the scalar spectral
index and tensor-to-scalar ratio
\begin{align}
  n_{\mathcal{R}} -1 & = \frac{ -n(n+ 2) + \alpha  \lambda e^{-\lambda  \phi } \phi ^{n+1} (2 \lambda  \phi -n)}{\phi ^2} \;, \label{nphi}  \\
  r & =  \frac{8 \left(n -\alpha  \lambda e^{-\lambda  \phi } \phi ^{n+1}\right)^2}{\phi ^2}  \label{rphi} \;,
\end{align}
which involve three model parameters $n$, $\l$ and $\a$ in the
slow-roll approximation. Hereafter, we restrict ourselves to
a quadratic potential, $n=2$, often considered the simplest
example for inflation~\cite{lin83}. The value of $\phi$ in~\eqref{nphi}
and \eqref{rphi} depends on the number of e-folds and the
value of $\phi_{\rm end}$ by setting max($\e_i,\d_i)(\phi_{\rm end})=1$.
For simplicity we set $N=60$.

\begin{figure}
\includegraphics[width=12cm,height=8cm]{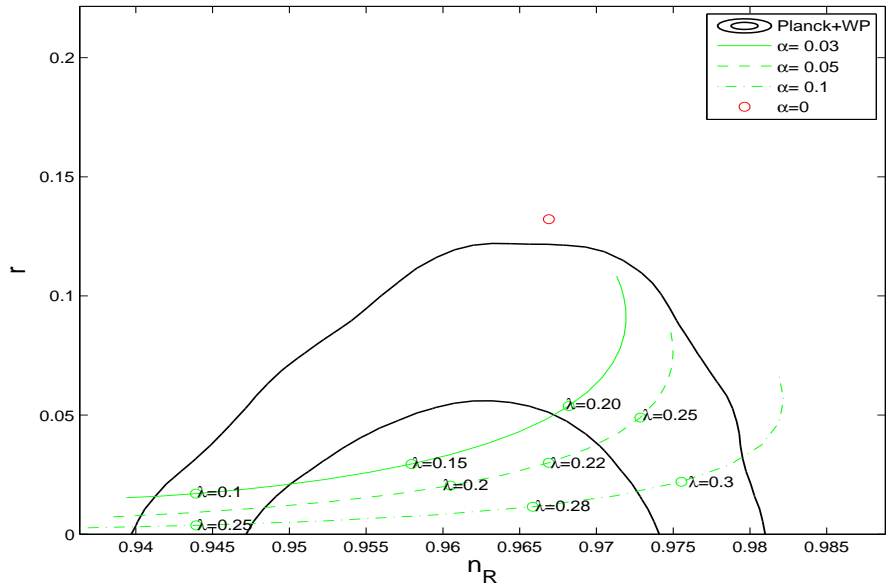}
\caption{Predicted $n_\RR$ versus $r$ in the model~\eqref{modelC}
with $n=2$ for different values of $\l$ and $\a$. Here we choose $N=60$.
The contours show the 68\% and 95\% CL from the Planck+WP data.}
\label{figC}
\end{figure}

In figure~\ref{figC} we plot the scalar spectral index and
tensor-to-scalar ratio for different values of $\l$ and $\a$.
There exist parameter regions in which the predicted $n_\RR$
and $r$ are excellently consistent with the Planck constraints.
We see that the scalar spectral index is sensitive to $\l$
for a given value of $\a$. Compared to the inverse monomial
coupling discussed in subsection B, these observable quantities
are more sensitive to the dilaton-like coupling.

\section{discussions and conclusions}
\label{sec4}

In this general slow-roll inflationary scenario, the potential
dominates the energy density of the Universe and the contribution
from the GB coupling is ignorable.
The GB coupling may slow down the evolution of the inflaton by
balancing the potential force, which decreases the energy scale
of the potential to be in agreement with the observed amplitude
of scalar perturbation. Hence, the tensor-to-scalar ratio is
suppressed. In principle, even for a steep potential slow-roll
inflation can occur with the help of the non-minimal coupling
of the inflaton to the GB term.
In the framework of the standard slow-roll inflation, it known
that the energy scale of inflation can be established by the
detection of the amplitude of tensor perturbations~\cite{ade13,guo11}.
In the presence of the GB coupling, one need further measure
the tensor tilt to establish the energy scale of inflation
because the new degree of freedom is introduced by the GB coupling.

Under the general slow-roll approximation, since the Hubble and GB
flow parameters, $\e_i$ and $\d_i$, are much smaller than 1,
the propagation speed of scalar perturbations~\eqref{c2RR} is very close to 1.
It is shown that the effect of the GB coupling on primordial non-Gaussianities
appears indirectly through the change of $c^2_\RR$~\cite{fel11}.
For the equilateral configuration, the non-linearity parameter is~\cite{fel11}
\bea
f_{\rm NL}^{\rm equil} \sim \frac{55}{36}\e_1 - \frac{25}{72}\d_1
 + \frac{5}{12}\frac{2\e_1\e_2-\d_1\d_2}{2\e_1-\d_1},
\ena
which means that these extra contributions from the GB coupling
remain of the order of small slow-roll parameters, just as in
the minimally-coupled single-field case. This is consistent with
the Planck's results~\cite{ade13n}.

In this paper we have applied the general slow-roll formalism to some specific
inflationary models in which the inflaton has a direct coupling
with the GB term. Since the consistency relation between the
tensor-to-scalar ratio and tensor spectral index is broken by the GB coupling,
we obtained a slightly tighter constraints on the tensor-to-scalar
ratio using the Planck+WP data when the tensor spectral index is
allowed to vary freely.
In the $r-n_\RR$ plane we then confront the models with observational
constraints. We found that there exit parameter regions in which
the predicted $r$ and $n_\RR$ are excellently consistent with the Planck constraints.
Moreover, in the scenario the non-linearity parameter is of the order of
slow-roll parameters, which is in agreement with the observational constraints.

\acknowledgments
This work is partially supported by the project of Knowledge
Innovation Program of Chinese Academy of Science,
NSFC under Grant No.11175225, and National Basic Research
Program of China under Grant No.2010CB832805.


\begin{thebibliography}{99}
\bibitem{hin12}
  G.~Hinshaw {\it et al.}  [WMAP Collaboration],
  arXiv:1212.5226.
\bibitem{ade13}
  P.~A.~R.~Ade {\it et al.}  [Planck Collaboration],
  arXiv:1303.5082.
\bibitem{cal85}
  C.~G.~Callan, Jr., E.~J.~Martinec, M.~J.~Perry and D.~Friedan,
  Nucl.\ Phys.\ B {\bf 262}, 593 (1985);
  D.~J.~Gross and J.~H.~Sloan,
  Nucl.\ Phys.\ B {\bf 291}, 41 (1987).
\bibitem{ant92}
  I.~Antoniadis, E.~Gava and K.~S.~Narain,
  Nucl.\ Phys.\ B {\bf 383}, 93 (1992)
  [arXiv:hep-th/9204030];
  I.~Antoniadis, J.~Rizos and K.~Tamvakis,
  Nucl.\ Phys.\ B {\bf 415}, 497 (1994)
  [arXiv:hep-th/9305025].
\bibitem{kaw98}
  S.~Kawai, M.~-a.~Sakagami and J.~Soda,
  Phys.\ Lett.\ B {\bf 437}, 284 (1998)
  [arXiv:gr-qc/9802033];
  S.~Kawai and J.~Soda,
  Phys.\ Lett.\ B {\bf 460}, 41 (1999)
  [arXiv:gr-qc/9903017].
\bibitem{top02}
  S.~Tsujikawa,
  Phys.\ Lett.\ B {\bf 526}, 179 (2002)
  [arXiv:gr-qc/0110124];
  A.~Toporensky and S.~Tsujikawa,
  Phys.\ Rev.\ D {\bf 65}, 123509 (2002)
  [arXiv:gr-qc/0202067];
  S.~Tsujikawa, R.~Brandenberger and F.~Finelli,
  Phys.\ Rev.\ D {\bf 66}, 083513 (2002)
  [arXiv:hep-th/0207228].
\bibitem{guo07a}
K.~Bamba, Z.~K.~Guo and N.~Ohta,
  Prog.\ Theor.\ Phys.\ {\bf 118}, 879 (2007)
  [arXiv:0707.4334];
K.~Andrew, B.~Bolen and C.~A.~Middleton,
  Gen.\ Rel.\ Grav.\ {\bf 39}, 2061 (2007)
  [arXiv:0708.0373];
R.~Chingangbam, M.~Sami, P.~V.~Tretyakov and A.~V.~Toporensky,
  Phys.\ Lett.\ B {\bf 661}, 162 (2008)
  [arXiv:0711.2122];
I.~V.~Kirnos and A.~N.~Makarenko,
  arXiv:0903.0083.
\bibitem{noj05}
S.~Nojiri, S.~D.~Odintsov and M.~Sasaki,
  Phys.\ Rev.\ D {\bf 71}, 123509 (2005)
  [arXiv:hep-th/0504052];
G.~Cognola, E.~Elizalde, S.~Nojiri, S.~D.~Odintsov and S.~Zerbini,
  Phys.\ Rev.\ D {\bf 73}, 084007 (2006)
  [arXiv:hep-th/0601008];
T.~Koivisto and D.~F.~Mota,
  Phys.\ Lett.\ B {\bf 644}, 104 (2007)
  [astro-ph/0606078];
S.~Tsujikawa and M.~Sami,
  JCAP {\bf 0701}, 006 (2007)
  [arXiv:hep-th/0608178];
T.~Koivisto and D.~F.~Mota,
  Phys.\ Rev.\ D {\bf 75}, 023518 (2007)
  [hep-th/0609155];
B.~ M.~Leith and I.~P.~Neupane,
  JCAP {\bf 0705}, 019 (2007)
  [arXiv:hep-th/0702002];
B.~C.~Paul and S.~Ghose,
  Gen.\ Rel.\ Grav.\ {\bf 42}, 795 (2010)
  [arXiv:0809.4131];
M.~R.~Setare and E.~N.~Saridakis,
  Phys.\ Lett.\ B {\bf 670}, 1 (2008)
  [arXiv:0810.3296];
J.~Sadeghi, M.~R.~Setare and A.~Banijamali,
  Phys.\ Lett.\ B {\bf 679}, 302 (2009)
  [arXiv:0905.1468];
J.~Sadeghi, M.~R.~Setare and A.~Banijamali,
  Eur.\ Phys.\ J.\ C {\bf 64}, 433 (2009)
  [arXiv:0906.0713];
P.~Wu and H.~Yu,
  Mod.\ Phys.\ Lett.\ A {\bf 25}, 2325 (2010);
J.~Moldenhauer, M.~Ishak, J.~Thompson and D.~A.~Easson,
  Phys.\ Rev.\ D {\bf 81}, 063514 (2010)
  [arXiv:1004.2459];
S.~Nojiri and S.~D.~Odintsov,
  Phys.\ Rept.\  {\bf 505}, 59 (2011)
  [arXiv:1011.0544];
M.~Iihoshi,
  Gen.\ Rel.\ Grav.\  {\bf 43}, 1571 (2011)
  [arXiv:1011.2088];
K.~Maeda, N.~Ohta and R.~Wakebe,
  Eur.\ Phys.\ J.\ C {\bf 72}, 1949 (2012)
  [arXiv:1111.3251].
\bibitem{bro05}
R.~A.~Brown, R.~Maartens, E.~Papantonopoulos and V.~Zamarias,
  JCAP {\bf 0511}, 008 (2005)
  [arXiv:gr-qc/0508116];
J.~H.~He, B.~Wang and E.~Papantonopoulos,
  Phys.\ Lett.\ B {\bf 654}, 133 (2007)
  [arXiv:0707.1180];
E.~N.~Saridakis,
  Phys.\ Lett.\ B {\bf 661}, 335 (2008)
  [arXiv:0712.3806].
\bibitem{hwa99}
  J.~-c.~Hwang and H.~Noh,
  Phys.\ Rev.\ D {\bf 61}, 043511 (2000)
  [arXiv:astro-ph/9909480];
  C.~Cartier, J.~-c.~Hwang and E.~J.~Copeland,
  Phys.\ Rev.\ D {\bf 64}, 103504 (2001)
  [arXiv:astro-ph/0106197];
  J.~-c.~Hwang and H.~Noh,
  Phys.\ Rev.\ D {\bf 71}, 063536 (2005)
  [arXiv:gr-qc/0412126].
\bibitem{sat08}
  M.~Satoh, S.~Kanno and J.~Soda,
  Phys.\ Rev.\ D {\bf 77}, 023526 (2008)
  [arXiv:0706.3585];
  M.~Satoh and J.~Soda,
  JCAP {\bf 0809}, 019 (2008)
  [arXiv:0806.4594];
  M.~Satoh,
  JCAP {\bf 1011}, 024 (2010)
  [arXiv:1008.2724];
  K.~Nozari and N.~Rashidi,
  arXiv:1310.3989.
\bibitem{guo07}
  Z.~K.~Guo, N.~Ohta and S.~Tsujikawa,
  Phys.\ Rev.\ D {\bf 75}, 023520 (2007)
  [arXiv:hep-th/0610336].
\bibitem{guo09}
  Z.~K.~Guo and D.~J.~Schwarz,
  Phys.\ Rev.\ D {\bf 80}, 063523 (2009)
  [arXiv:0907.0427].
\bibitem{guo10}
  Z.~K.~Guo and D.~J.~Schwarz,
  Phys.\ Rev.\ D {\bf 81}, 123520 (2010)
  [arXiv:1001.1897].
\bibitem{sch01}
D.~J.~Schwarz, C.~A.~Terrero-Escalante and A.~A.~Carcia,
  Phys.\ Lett.\ B {\bf 517}, 243 (2001)
  [arXiv:astro-ph/0106020];
S.~M.~Leach, A.~R.~Liddle, J.~Martin and D.~J.~Schwarz,
  Phys.\ Rev.\ D {\bf 66}, 023515 (2002)
  [arXiv:astro-ph/0202094];
D.~J.~Schwarz and C.~A.~Terrero-Escalante,
  JCAP {\bf 0408}, 003 (2004)
  [arXiv:hep-ph/0403129].
\bibitem{che12}
  C.~Cheng, Q.~G.~Huang, X.~D.~Li and Y.~Z.~Ma,
  Phys.\ Rev.\ D {\bf 86}, 123512 (2012)
  [arXiv:1207.6113].
\bibitem{lin83}
  A.~D.~Linde,
  Phys.\ Lett.\ B {\bf 129}, 177 (1983).
\bibitem{guo11}
A.~R.~Liddle,
  Phys.\ Rev.\ D {\bf 49}, 739 (1994)
  [astro-ph/9307020];
Z.~K.~Guo, D.~J.~Schwarz and Y.~Z.~Zhang,
  Phys.\ Rev.\ D {\bf 83}, 083522 (2011)
  [arXiv:1008.5258];
J.~Martin, C.~Ringeval and V.~Vennin,
  arXiv:1303.3787;
S.~Choudhury and A.~Mazumdar,
  arXiv:1306.4496.
\bibitem{fel11}
  A.~De Felice and S.~Tsujikawa,
  JCAP {\bf 1104}, 029 (2011)
  [arXiv:1103.1172].
\bibitem{ade13n}
  P.~A.~R.~Ade {\it et al.}  [Planck Collaboration],
  arXiv:1303.5084.
\end{thebibliography}
\end{document}